\documentclass[epj,twocolumn]{webofc}
\usepackage[varg]{txfonts}   
\woctitle{CGS15}

\begin{document}

\title{The structure of rotational bands in alpha-cluster nuclei}

\author{Roelof Bijker\inst{1}\fnsep\thanks{\email{bijker@nucleares.unam.mx}} 
}

\institute{ICN-UNAM, AP 70-543, 04510 Mexico DF, Mexico}

\abstract{In this contribution, I discuss an algebraic treatment of alpha-cluster nuclei based 
on the introduction of a spectrum generating algebra for the relative motion of the alpha-clusters. 
Particular attention is paid to the discrete symmetry of the geometric arrangement of the 
$\alpha$-particles, and the consequences for the structure of the rotational bands in the $^{12}$C 
and $^{16}$O nuclei.}

\maketitle

\section{Introduction}
\label{intro}

\begin{figure}
\centering
\includegraphics[width=8cm,clip]{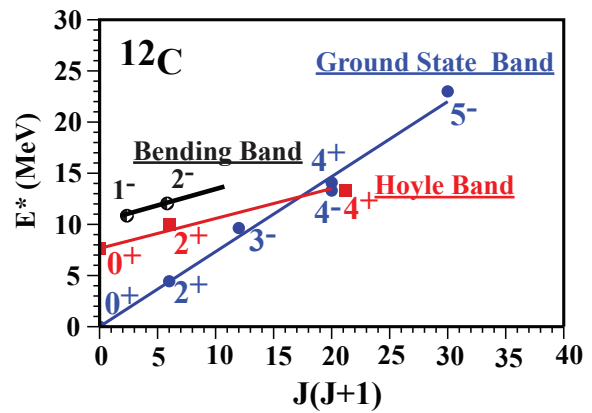}
\caption{Rotational $\alpha$-cluster states of the ground-state band, the Hoyle band and
the bending vibration in $^{12}$C \cite{C12}.}
\label{bands}
\end{figure}

Recently, there has been a lot of renewed interest in the structure of $^{12}$C \cite{FreerFynbo}. 
The experimental identification of new rotational states of the ground state band \cite{Fourminus,C12} 
and the Hoyle band \cite{Twoplus,Fourplus} (see Fig.~\ref{bands}) has stimulated a large theoretical 
effort to understand the underlying geometric structure of $^{12}$C. Among others there are studies based 
on antisymmetrized molecular dynamics \cite{AMD}, fermionic molecular dynamics (FMD) \cite{FMD},   
BEC-like cluster model \cite{BEC}, {\it ab initio} no-core shell model \cite{NCSM}, lattice EFT \cite{EFT}, 
no-core symplectic model \cite{Draayer} and the Algebraic Cluster Model (ACM) \cite{BI,C12}. 
Earlier work on $\alpha$-cluster models includes studies by Wheeler \cite{Wheeler}, Brink \cite{Brink} 
and Robson \cite{Robson}. For a recent review on the structure of $^{12}$C, see \cite{FreerFynbo}. 

In this contribution, the spectroscopy of $\alpha$-cluster states in $^{12}$C and $^{16}$O is analyzed 
in the framework of the Algebraic Cluster Model. It is discussed how the structure of rotational bands 
can be used to determine the underlying geometric arrangement of the $\alpha$-clusters and hence to 
distinguish between different theoretical approaches of $\alpha$-cluster nuclei. 

\section{Algebraic Cluster Model}
\label{sec-1}

The Algebraic Cluster Model (ACM) is an interacting boson model to describe the relative 
motion of the $n$ clusters based on the spectrum generating algebra of $U(\nu+1)$ where 
$\nu=3(n-1)$ represents the number of relative spatial degrees of freedom. 
For the two-body problem the ACM reduces to the $U(4)$ vibron model 
\cite{vibron}, while for three-body clusters it leads to the $U(7)$ model \cite{BI,C12,BIL}. 
and for four-body clusters to the $U(10)$ model \cite{RB,O16}. 

The spatial degrees of freedom for $n$-body systems are taken as the relative Jacobi 
coordinates $\vec{\rho}_k$ 
\begin{eqnarray*}
\vec{\rho}_{k} = \frac{1}{\sqrt{k(k+1)}} 
\left( \sum_{i=1}^{k} \vec{r}_{i} - k \vec{r}_{k+1} \right) ~, 
\end{eqnarray*}
with $k=1,\ldots,n-1$, and their conjugate momenta. 
Here $\vec{r}_i$ are the coordinates of the $i$-th cluster 
($i=1,\ldots,n$). Instead of a formulation in terms of coordinates and momenta the method of 
bosonic quantization is used which consists of introducing a dipole boson 
($b^{\dagger}_{k}$) for each independent relative Jacobi vector, 
and an auxiliary scalar boson ($s^{\dagger}$). The scalar boson does not 
represent an independent degree of freedom, but is added under the restriction that the total 
number of bosons $N=n_s+\sum_k n_k$ is conserved. This procedure leads to a compact 
spectrum generating algebra of $U(3n-2)$ whose model space is spanned by the symmetric 
irreducible representation $[N]$ which contains the oscillator shells with 
$n_b=\sum_k n_{k}=0,1,2,\ldots, N$. The introduction of the scalar boson makes it possible 
to investigate the dynamics of $n-1$ vector degrees of freedom including situations in which 
there is a mixing of oscillator shells. 

In the application to $\alpha$-cluster nuclei the Hamiltonian has to be invariant under the 
permuation group $S_n$ for $n$ identical clusters. For the case of the harmonic oscillator 
the construction of states with good permutation symmetry was studied by Kramer and 
Moshinsky \cite{KM}. However, since in the ACM the number of oscillator shells can be large 
and oscillator shells may be mixed, the wave functions with good permutation symmetry are 
generated numerically, and their permutation symmetry is determined by considering their 
properties under the interchange of the first two clusters $P(12)$ and the cyclic permutation 
$P(12 \cdots n)$ \cite{QTS}.  

As a result, the ACM wave functions are characterized by the total number of bosons $N$, 
the angular momentum and parity $L^P$ and permutation symmetry $t$. In the case of (identical) 
$\alpha$-particles, the wave functions have to be completely symmetric.  

\subsection{The Nucleus $^{12}$C}

As a first application, the spectroscopic properties of $^{12}$C are analyzed in the oblate top 
limit of the ACM for three-body clusters which corresponds to the geometric configuration of 
three $\alpha$-particles located at the vertices of an equilateral triangle \cite{BI,C12} 
with ${\cal D}_{3h}$ symmetry. The energy spectrum consists of a series of rotational 
bands labeled by $(v_1,v_2^{\ell_2})$, where $v_1$ corresponds to the breathing vibration with 
$A$ symmetry and $v_2$ to the doubly degenerate bending vibration with $E$ symmetry;
$\ell_2$ denotes the vibrational angular momentum of the doubly degenerate vibration, 
The states are further labeled by the angular momentum $L$, its projection on the symmetry axis $K$, 
and the parity $P$.

The structure of rotational bands depends on the discrete symmetry of the vibrations. 
For vibrational bands with $(v_1,0^{0})$ and $A$ symmetry, the allowed values of the angular 
momenta and parity are $L^P=0^+$, $2^+$, $4^+$, $\ldots$, with $K=0$ and $L^P=3^-$, $4^-$,
$5^-$, $\ldots$, with $K=3$. The three-fold symmetry excludes states with $K=1$ and $K=2$
and leads to the prediction of the existence of a $L^P = 4^\pm$ parity doublet in the 
$(v_1,0^{0})$ vibrational band both for the ground band ($v_1=0$) and the Hoyle band ($v_1=1$). 
For the bending vibration with $(0,1^{1})$ and $E$ symmetry, the rotational
sequence is given by $L^{P}=1^-$, $2^-$, $3^-$, $4^-$, $\ldots$, with $K=1$,
$L^{P}=2^+$, $3^+$, $4^+$, $\ldots$, with $K=2$ and $L^{P}=4^{+}$, $\ldots$, with $K=4$.
Since in the application to the cluster states of $^{12}$C, the vibrational and rotational
energies are of the same order, one expects large rotation-vibration couplings and 
anharmonicities. The large anharmonicities lead to an increase of the rms radius of the 
vibrational excitations relative to that of the ground state. 

\begin{figure}
\centering
\includegraphics[width=8cm,clip]{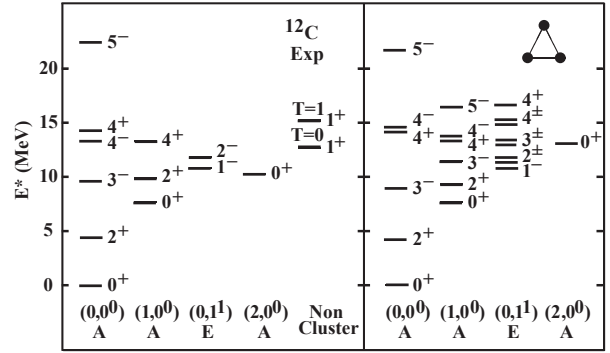}
\caption{Comparison between the low-lying experimental spectrum of $^{12}$C and the energies
of the oblate top \cite{C12}. The levels are organized in columns corresponding to the ground state band 
and the vibrational bands with $A$ and $E$ symmetry of an oblate top with triangular symmetry. 
The last column on the left-hand side, shows the lowest observed non-cluster ($1^+$) levels.}
\label{carbon}
\end{figure}

The energy eigenvalues of the oblate top, up to
terms quadratic in the rotation-vibration interaction, are given by 
\begin{eqnarray*}
E = E_0 + \omega_{1}(v_{1}+\frac{1}{2}) \left( 1-\frac{v_1+1/2}{N} \right) \hspace{1.5cm}
\nonumber\\
+ \omega_{2}(v_{2}+1) \left( 1-\frac{v_2+1}{N+1/2} \right)
+ \kappa_{2} \, (K \mp 2\ell_{2})^{2}
\nonumber\\
+ \left[\kappa_1 + \lambda_{1} \, (v_{1}+\frac{1}{2}) + \lambda_{2} \, (v_{2}+1) \right] L(L+1) ~.
\end{eqnarray*}
This formula includes both the anharmonicities which depend on $N$ and the vibrational
dependence of the moments of inertia \cite{C12}. In Fig.~\ref{carbon} we show a comparison of the
cluster states of $^{12}$C with the spectrum of the oblate top. 

For the ground state band of $^{12}$C both the positive and negative parity states have been 
observed, including a nearly degenerate doublet of states with $L^P=4^{\pm}$, whereas for the 
Hoyle band only those with positive parity have been identified. An interesting open question 
concerns the geometric structure of the Hoyle band in $^{12}$C. In order to distinguish between 
a bent-arm configuration as suggested in lattice EFT calculations \cite{EFT}, and an equilateral 
triangular configuration as in the ACM \cite{C12}, the identification of the negative parity 
$3^-$ and $4^-$ states of the Hoyle band is crucial. The selectivity of $\gamma$-ray beams 
as well as electron beams could help to populate the states of 
interest and resolve the broad interfering states. 
The $4^-$ Hoyle state which is predicted to be nearly degenerate with the $4^+$ Hoyle state, 
can be measured for example in 180$^\circ$ electron scattering off $^{12}$C \cite{Darmstadt}. 
We note that a (broad) $3^-$ state was suggested to lie between 11 and 14 MeV \cite{Fourminus}
which is close to the predicted energy shown in Fig.~\ref{carbon}. 

\subsection{The Nucleus $^{16}$O}

A similar analysis of $^{16}$O in the ACM for four identical alpha-particles gives rises to the 
rotation-vibration spectrum of a spherical top with tetrahedral symmetry which corresponds to a 
geometric configuration of four alpha-particles located at the vertices of a regular tetrahedron. 
The vibrational spectrum is labeled by $(v_1,v_2,v_3)$, where $v_1$ corresponds to the breathing 
vibration with $A$ symmetry, $v_2$ represents a two-dimensional bending vibration with $E$ symmetry 
and $v_3$ a three-dimensional bending vibration with $F$ symmetry. The ground state band and the 
breathing vibration consist of states with angular momentum and parity $L^P=0^+$, $3^-$, $4^+$, 
$6^{\pm}$, $\ldots$, while vibrational excitations with $E$ symmetry have 
$L^P=2^{\pm}$, $4^{\pm}$, $5^{\pm}$, $\ldots$, and those with $F$ symmetry 
$L^P=1^-$, $2^+$, $3^{\pm}$, $4^{\pm}$, $\ldots$ \cite{O16}.

\begin{figure}
\centering
\includegraphics[width=5cm,clip]{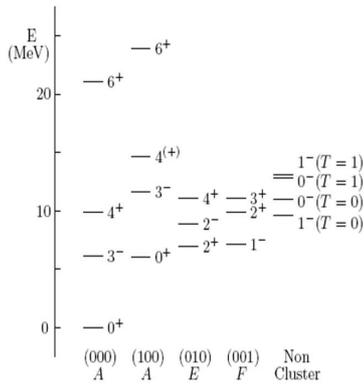}
\caption{The observed spectrum of $^{16}$O. The levels are organized in columns 
corresponding to the ground state band and the three vibrational bands with $A$, $E$ and $F$ 
symmetry of a spherical top with tetrahedral symmetry \cite{O16}. The last column shows 
the lowest non-cluster levels.} 
\label{oxygen}
\end{figure}

Fig.~\ref{oxygen} shows that these rotational sequences are indeed observed in the experimental 
spectrum of $^{16}$O. 

\section{Conclusions}

In conclusion, I discussed the ACM to describe the dynamics of $\alpha$-cluster nuclei with $A=4n$ 
nucleons. In an application to $^{12}$C, the alpha-particles are located at the vertices of an 
equilateral triangle with ${\cal D}_{3h}$ symmetry \cite{BI,C12}, and in $^{16}$O at the vertices 
of a regular tetrahedron with ${\cal T}_d$ symmetry \cite{O16}. The rotational sequences consist of both 
positive and negative parity states, just as observed experimentally in $^{12}$C \cite{C12} and 
$^{16}$O \cite{O16}, and can be considered fingerprints of the underlying discrete symmetry of the 
geometric configuration of the alpha-particles. Even though the present contribution is based on a 
study of the energy spectrum of $\alpha$-cluster nuclei, it is important to note that an analyisis of 
electromagnetic transition rates, form factors and $B(EL)$ values supports the evidence for triangular 
symmetry in $^{12}$C \cite{BI,C12} and tetrahedral symmetry in $^{16}$O \cite{O16}. 
Other signatures of $\alpha$-cluster configurations may be found in relativistic nuclear collisions 
\cite{rnc} and giant dipole resonances \cite{gdr}.

It is hoped that the results of Refs.~\cite{C12} and \cite{O16}, reviewed here, will stimulate 
a dedicated experimental research program on the structure of $\alpha$-cluster nuclei in order to shed 
new light on the clustering phenomena in light nuclei, in particular on the structure of the Hoyle band. 
 
\section*{Acknowledgements}

The author wishes to thank Franco Iachello and Moshe Gai for interesting discussions.  
This work was supported in part by PAPIIT project IN 107314.

\end{document}